\documentclass[useAMS,usenatbib]{mn2e}
\usepackage{epsfig}
\usepackage{longtable}
\usepackage{natbib}

\newdimen\figurewidth
\def\sms{SMS}
\def\mr{\mathrm}
\def\mr{\mathrm}
\def\ledd{{\cal L}_{\mr{Edd}}}

\def\func{{\cal W}}

\def\gtrsim{\mathrel{\hbox{\rlap{\hbox{\lower3pt\hbox{$\sim$}}}\raise2pt\hbox{$>$}}}}
\def\lesssim{\mathrel{\hbox{\rlap{\hbox{\lower3pt\hbox{$\sim$}}}\raise2pt\hbox{$<$}}}}

\makeatletter

\newcommand{\Rmnum}[1]{\expandafter\@slowromancap\romannumeral #1@}
\makeatother

\title[The Super-Eddington Nature of Super-Massive Stars]{The Super-Eddington Nature of Super-Massive Stars}
\author[Dotan \& Shaviv]{Calanit Dotan$^{1}$ and Nir J. Shaviv$^{1}$\\
\noindent
$^{1}$Racah Institute of Physics, Hebrew University of Jerusalem, Jerusalem 91904, Israel }

\usepackage{times}

\usepackage{colordvi}

\RequirePackage[colorlinks=true
,urlcolor=blue
,anchorcolor=blue
,citecolor=blue
,filecolor=blue
,linkcolor=blue
,menucolor=blue
,pagecolor=blue
,linktocpage=true
,pdfproducer=medialab
]{hyperref}

\begin{document}

\pagerange{\pageref{firstpage}--\pageref{lastpage}} \pubyear{2009}

\maketitle

\label{firstpage}

\begin{abstract}
Supermassive stars (SMS) are massive hydrogen objects, that slowly radiate their gravitational binding energy. Such hypothetical primordial objects may have been the seed of the massive black holes (BHs) observed at the centre of galaxies. Under the standard picture, these objects can be approximately described as $n=3$ polytropes, and they are expected to shine extremely close to their Eddington luminosity. Once however, one considers the porosity induced by instabilities near the Eddington limit, which give rise to super-Eddington states, the standard picture should be modified. We study the structure, evolution and mass loss of these objects. We find the following. First, the evolution of SMSs is hastened due to their increased energy release. They accelerate continuum driven winds. If there is no rotational stabilization, these winds are insufficient to ``evaporate" the objects, such that they can collapse to form supermassive BHs, however, they do prevent SMSs from emitting a copious amount of ionizing radiation.  If the SMSs are rotationally stabilized, the winds ``evaporate" the objects until a normal sub-Eddington star remains, having a mass of a few 100$M_\odot$. 
\end{abstract}


\section{Introduction}
Super-massive stars (henceforth \sms s) are hydrogen objects  with large masses, in the range of $10^4-10^8M_{\odot}$. Their state is described as a hydrostatic equilibrium between their self gravity on one hand, and radiation pressure on the other.  Since radiation leaks out as the objects shine, the thermal energy is replenished through the slow release of the gravitational binding energy. This is unlike the less massive stars, where the central temperatures and densities are high enough to ignite thermonuclear reactions, the energy of which sustains the hydrostatic equilibrium. Thus, {\sms}s look like massive radiatively supported stars, but without nuclear reactions \citep{Wagoner1969,WeinbergBook,ShapiroBook}. 



Under the standard picture (summarized in \S\ref{sec:std_pic}), SMSs are expected to shine at nearly the Eddington limit. However, it is now known that objects shining near this luminosity should develop instabilities, reduce the effective opacity in their atmosphere and attain super-Eddington luminosities  \citep{ShavivInstabilities,ShavivNovae}. SMSs are expected to be no different, and should be described with super-Eddington states. 

The main qualitative difference introduced by these super-Eddington states, is the acceleration of a continuum driven wind. This high mass loss rate may affect their evolution. It is also responsible for a larger photosphere which may affect the amount of ionizing radiation emitted by the SMSs. These effects are studied in the present work.

Two further complications should be considered. First, it was demonstrated by \cite{Fowler66} that a dynamically insignificant amount of rotation (namely, rotation for which the star is still close to being spherical) can stabilize the SMS against the GR collapse into a massive black hole. In the present analysis, we will consider both cases---with and without rotational stabilization.

The second complication was proposed by \cite{Begelman2010}, who considered a more realistic model for the formation of SMSs. It was argued that under a natural scenario, SMSs can form with an entropy inversion, such that the outer layers which are accreted later, have a higher entropy. Such SMSs are not $n=3$ polytropes anymore. Instead, they have a small convective core surrounded by a massive convectively stable envelope. We will not consider this type of model here. Just as the standard model is modified here to consider the super-Eddington states, Begelman's scenario should be likewise modified. We defer this to subsequent work. 

\subsection{Super Massive Stars: The Standard Picture}
\label{sec:std_pic}
To first approximation, SMSs can be modeled as pure polytropes \citep[e.g.,][]{ShapiroBook}. That is, they can be approximated using $P=K\rho^\Gamma$. Since radiation pressure dominates the gas pressure in SMSs, the total pressure is approximately given by the radiation pressure. Together with equation of hydrostatic equilibrium,
\begin{equation}
\frac{dP}{dr}=-\frac{GM(r)\rho}{r^2}
\label{eq:hydro1}
\end{equation}
and equation of radiative transfer 
\begin{equation}
L=-4\pi r^2\frac{c}{3\kappa\rho}\frac{d}{dr}\left(aT^4\right),
\label{eq:radiation1}
\end{equation}
it is easy to see that the star radiates at its Eddington limit,
\begin{equation}
L=\ledd=\frac{4\pi cGM}{\kappa}.
\label{eq:Ledd}
\end{equation}
Once we will allow the opacity to decrease in the super-Eddington states, we see that the object's luminosity will necessarily increase. 
\subsubsection{Stability of non-rotating SMSs}
\label{sec:stability}
One of the most interesting aspects of non-rotating SMSs, is their instability towards gravitational collapse, which arises from corrections due to General Relativity. We succinctly summarize the conditions under which this instability arises (for more details see \citealt{ShapiroBook}). These are important as they will be required once we will describe the evolution of the SMSs.
The Newtonian energy of a polytrope is given by
\begin{equation}
E=\int_0^R 4\pi r^2 \left(u+\frac{GM\rho}{r^2}\right)dr,
\label{eq:Epoly1}
\end{equation}
where $u$ is the internal energy density, $u=P/(\Gamma-1)$ and $\Gamma=4/3$ is the polytropic index. This integral is solved for an $n=3$ polytrope to get
\begin{equation}
E=k_1K\rho_c^{1/3} M-k_2G\rho_c^{1/3}M^{5/3},
\label{eq:Epoly2}
\end{equation} 
where $k_1$ and $k_2$ are constants, and $\rho_c$ is the central density of the SMS. For a given mass, the equilibrium configuration occurs when $\partial{E}/\partial{\rho_c}=0$. This condition of equilibrium provides a relation between the total mass and the specific entropy of radiation,
\begin{equation}
s_\mr{rad}=\frac{4m_Ha}{3}\left(\frac{3\pi G}{a}\right)^{3/4}\sqrt\frac{M}{8\pi}.
\label{eq:srad1}
\end{equation}
Higher order corrections to the energy can be obtained by adding the effects of gas pressure to the internal energy and the post-Newtonian (i.e., GR) corrections to the gravitational energy. The total energy of the star including these corrections is \citep[chap. 6]{ShapiroBook}
\begin{equation}
E=AM\rho_c^{1/3}-BM^{5/3}\rho_c^{1/3}+CM\rho_c^{1/3}ln\rho_c-DM^{7/3}\rho_c^{2/3},
\label{eq:Egr}
\end{equation}
where $A$, $B$, $C$ and $D$ are constants, (the terms proportional to $C$ and $D$ are derived from the additional effects of gas pressure and post-Newtonian corrections). The condition for equilibrium configuration $\partial{E}/\partial{\rho_c}=0$, gives the central density as a function of the mass and the specific entropy. 
General relativity instabilities appears when $\partial^2E/\partial\rho_c^2=0$. Differentiating eq.\ (\ref{eq:Egr}) twice, yields:
\begin{equation}
0=\frac{1}{2}CM\rho_c^{-2/3}-\frac{1}{3}DM^{7/3}\rho_c^{-1/3}.
\label{eq:rho_crit}
\end{equation}
Thus, there is a critical value for the density, above which the star becomes unstable. This analysis will not change once we introduce the existence of super-Eddington states.  

\subsubsection{Stabilization by rotation}
\label{sec:Rotation}
The General Relativistic instability disappears if the SMS is rotating fast enough \citep{Fowler66}. In this case, the Newtonian term for the rotational energy should be added to the total energy (eq. \ref{eq:Egr}). It is given by
\begin{equation}
\Psi=\frac{1}{2}\int r^2 \omega^2 sin^2(\theta) \rho dV,
\label{eq:Psi}
\end{equation}
where $\omega$ is the angular velocity and $\theta$ is the polar angle measured from the axis of rotation. 
For a constant\footnote{For the more general case of non-constant angular momentum see \cite{Fowler66}.} angular momentum $\Phi$, the critical value of $\Phi$ which stabilize the SMS against GR instability is given by
\begin{equation}
\frac{\Phi_{cr}}{M}=3.6\times 10^{15}\frac{M}{M_{\odot}} cm^2 sec^{-1},
\label{eq:Phi_crit}
\end{equation}
this value is relatively small even for a SMS with $M=10^8M_{\odot}$.
If $\Phi>\Phi_{cr}$, rotation  prevents the gravitational collapse. 




\section{Background: Super-Eddington States}

Before proceeding to construct super-Eddington SMSr models, we begin by reviewing the relevant physics pertaining the emergence of super-Eddington states. These include three particular elements. First, the rise of inhomogeneities was shown \citep{ShavivPorous} to reduce the effective opacity. Second, once a super-Eddington state arises, strong continuum driven winds are accelerated. Last, if the wind mass loss is too large, wind stagnation and a photon-tired state arises \citep{PhotonTired}. In it, a layer is formed in which strong shocks mediate a high energy flux without an excessive  mass flux. These components are necessary building blocks for the super-Eddington states, and we therefore review them below. Two more examples where this theory is applied can be found in \cite{DotanDisks} and \cite{DotanHalos}.

\subsection{The rise of super-Eddington states}
\label{sec:SEDstate}
According to common wisdom, objects cannot shine beyond their classical
Eddington limit, $\ledd$, since no hydrostatic solution exists.  In
other words, if objects do pass $\ledd$, they are highly dynamic.
They have no steady state, and a huge mass loss should occur since their
atmospheres are then gravitationally unbound and they should therefore
be expelled.  Thus, astrophysical objects according to this picture, can
pass $\ledd$ but only for a short duration corresponding to the time
it takes them to dynamically stabilize once super-Eddington conditions are forced.

For example, this can be  seen in detailed 1D numerical simulations of nova thermonuclear run aways, where novae can be in a super-Eddington state but only for several dynamical time scales \citep[e.g.,][]{Prialnik1992}.
However, once they do stabilize, they are expected and indeed do reach
 in the simulations a sub-Eddington state.  Namely, we
naively expect to find no steady state super-Eddington
atmospheres. This, however, is not the case in nature, where nova eruptions are clearly super-Eddington for durations which are orders of magnitude longer then their dynamical time scale \citep{ShavivNovae}. This is exemplified with another clear super-Eddington object---the great eruption of the massive star $\eta$-Carinae, which was a few times super-Eddington for over 20 years \citep{ShavivEta}

The existence of a super-Eddington state can be naturally explained, once we consider the following.
\begin{enumerate}
\item Atmospheres become unstable as they approach the Eddington limit. In addition to instabilities that operate under various special conditions (e.g., Photon bubbles in strong magnetic fields, \citealt{Bubbles1,Bubbles2,Bubbles3}, or s-mode instability under special opacity laws, \citealt{Opacity1,Opacity2}), two instabilities operate in Thomson scattering atmospheres, \citep{ShavivInstabilities}. It implies that {\em all atmospheres will become unstable already before reaching the Eddington limit}.
\item The effective opacity for calculating the radiative force on an inhomogeneous atmosphere is not necessarily the microscopic opacity. Instead, it is given by
\begin{equation}
\label{eq:effectiveOpacity}
    \kappa_{V}^{\mr{eff}} \equiv {\left\langle F\kappa_{V}\right\rangle_{V}
      \over  \left\langle F \right\rangle_{V}},  
\end{equation}
where $\left< ~\right>_V$ denotes volume averaging and F is the flux.
 The situation is very similar to the Rosseland vs. Force opacity means used in non-gray atmospheres, where the inhomogeneities are in frequency space as opposed to real space. For the special case of Thomson scattering, the effective opacity is always reduced.

\end{enumerate}

Thus, we find that as atmospheres approach their classical Eddington limit, they will necessarily become inhomogeneous. These inhomogeneities will necessarily reduce their effective opacity such that the effective Eddington limit will not be surpassed even though the luminosity can be super-classical-Eddington. This takes place in the external part of luminous objects, where the radiation diffusion time scale is shorter than the dynamical time scale in the atmosphere. Further inside the atmosphere, convection is necessarily excited such that the total energy flux may be super-Eddington, but the radiative part of it is necessarily sub-Eddington with the convective flux carrying the excess \citep{Joss1973}. 

\subsection{Super-Eddington Winds}
\label{sec:SEDwinds}

The atmosphere remains sub-Eddington while being classically super-Eddington only as long as the inhomogeneities comprising the atmosphere are optically thick. This condition will break at some point where the density is low enough. At this height, the effective opacity returns to its microscopic value and hence the radiative force becomes super-Eddington again.  Above this point we obtain a thick wind. Because it is optically thick, the conditions at the wind affect the structure at its base.

At the critical point, the radiative and gravitational forces balance each other. This point will coincide with a sonic surface (where the mass loss velocity equals the local speed of sound) for a steady state wind. This allows us to obtain the local mass loss rate which is given by
\begin{equation}
\dot{m}_{\mr{w}}=4\pi r_\mr{s}^2\rho_{\mr{crit}}v_s({r_{\mr{s}}})=const.
\label{eq:mdot}
\end{equation}
where $r_{\mr{s}}$ is the critical (sonic) radius, $\rho_{\mr{crit}}$ is the density at this radius and $v_s$ is the local isothermal speed of sound.

Based on the fact that instabilities develop structure with a typical size comparable to the density scale height in the atmosphere, it is possible to estimate the average density at the sonic point \citep{ShavivNovae}. Using this density the mass loss can be estimate to be
\begin{equation}
\dot{m}_\mr{w}={\cal W}\frac{L-\ledd}{cv_s}.
\label{eq:mdot2}
\end{equation}
where $\func$ is a dimensionless wind ``function''.  In principle, $\func$
can be calculated from first principles only after the nonlinear state
of the inhomogeneities is understood.  This however is still lacking
as it requires elaborate 3D numerical simulations of the nonlinear
steady state.  Nevertheless, it can be done in several
phenomenological models which depend on geometrical parameters such as the average size of the inhomogeneities in units of the
scale height ($\beta \equiv d / l_{p}$), the average ratio between the surface area and volume of
the blobs in units of the blob size ($\Xi$), and the volume filling
factor $\alpha$ of the dense blobs.
For example, in the limit in which the blobs
are optically thick, one can show that
$
\func = {3 \Xi / 32 \sqrt{\nu} \alpha \beta (1-\alpha)^{2}}
$ \citep{ShavivNovae},
with $\nu$ being the ratio between the effective speed of sound in the
atmosphere to the adiabatic one. Thus, $\func$ depends only on
geometrical factors. It does not depend explicitly on the Eddington
parameter $\Gamma \equiv L/\ledd $.  It yields typical values of $\func \sim 1-10$.

\section{The Model}
We now proceed to describe the underlying physics necessary for the construction of a SED SMS. This can be divided into three parts. The first is a description of the internal structure of the SMS. Although it can be approximately described by a polytrope, we will do so with the more complete set of equations. The second part is that of the porous atmosphere which forms the outer static layer of the object. This layer is responsible for the reduced opacity which allows for the emergence of the super-Eddington state. Above this layer, we find the third part, which is that of a continuum driven wind. Because this wind is optically thick, it affects the structure, and later the appearance of the SMSs.

\begin{figure}
\centerline{\epsfig{file=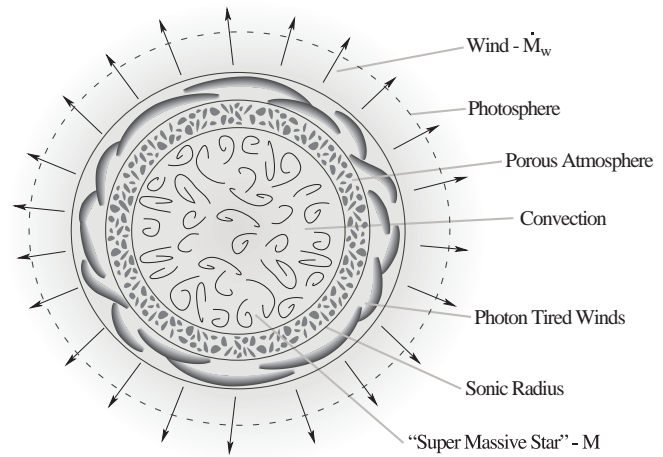,width=3.5in}}
\caption{ The general structure of a SMS
above the Eddington limit. In regions where the density is high enough, the Eddington luminosity cannot be surpassed. Deep inside the star, convection will necessarily arise as the Eddington luminosity is approached, keeping the luminosity below $\ledd$. Further out, it was discovered that instabilities will arise as $L$ approaches $L_{Edd}$ \citep[e.g.][]{ShavivInstabilities}, and that the ensuing nonlinear structure will have a reduced effective opacity \citep{ShavivPorous}.  However, as the overall density declines with height, the clumps become optically thin and so are exposed to the full radiative flux,
thereby increasing the net force above gravity and introducing a net supersonic
outflow, forming a continuum-driven stellar wind.}
\label{fig:Cartoon}
\end{figure}

The general structure is summarized in fig.\ \ref{fig:Cartoon}, and further elaborate below.

\subsection{The Stellar Structure Equations}
\subsubsection{Convective Core}
The convective zone begins at the center of the star and occupies the bulk of the hydrostatic radius and most of the SMS mass. In this region, the first two equations describing the stellar structure (in radial coordinates, using standard notation), are the equation of hydrostatic equilibrium
\begin{equation}
\frac{dP}{dr}=-\frac{Gm(r)\rho}{r^2},
\label{eq:hydrostatic_eq}
\end{equation}
and the equation of mass conservation,
\begin{equation}
\frac{dm}{dr}=4{\pi}r^2\rho.
\label{eq:mass_conser}
\end{equation}
The third is the equation of state, which is obtained as a combination of an ideal ionized gas and black body radiation,
\begin{equation}
P=P_r+P_g=\frac{1}{3}aT^4+\frac{{\rho}kT}{{\mu}m_p},
\label{eq:pressure}
\end{equation}
where $\mu=(2 X + 1.5 Y)^{-1}$ is the mean molecular weight, and $X$ \& $Y$ are the hydrogen and helium mass fractions respectively. 

The last equation describes energy transfer inside the star. Under conditions prevailing in SMSs, the relevant energy transfer takes place through {\em convection}. The temperature gradient is hence given by
\begin{equation}
\frac{dT}{dr}=\frac{\gamma-1}{\gamma}\frac{T}{P}\frac{dP}{dT},
\label{eq:gradT}
\end{equation}
where $\gamma$ is the adiabatic index.
Convection is present if the standard Schwarzschild criterion is satisfied, but although this condition is satisfied all the way out to the stellar surface, at a certain radius the energy transport becomes purely radiative, as convection becomes inefficient. For convection to be efficient, the convective flux must be smaller than the maximum possible, which is given by
\begin{equation}
L_\mr{conv,max}=4\pi r^2\rho v_s^3,
\label{eq:Lconv}
\end{equation}
where $v_s$ is the adiabatic speed of sound. 
\subsubsection{Atmosphere}
If the density is too low, convection becomes inefficient, the energy transfer takes place through radiation diffusion, and the radiative luminosity, $L_\mr{rad}$ becomes super-Eddington.
The equation of radiative transfer is modified to take into account the porous nature of the atmosphere:
\begin{equation}
\frac{dT}{dr}=-\frac{3\kappa_\mr{eff}L\rho}{16\pi acr^2T^3},
\label{eq:dT_atm}
\end{equation} 
where $\kappa_\mr{eff}$ is the effective opacity. As described in \S\ref{sec:SEDstate}, when the flux approaches a critical value, the atmosphere develops inhomogeneities and the gas becomes porous, such that the radiative force exerted on the gas is reduced. 
We assume that the relation between the effective Eddington factor $\Gamma_{\mr{eff}} \equiv L/L_\mr{eff}$ and the classical Eddington factor $\Gamma \equiv L/\ledd$ is empirically given by
\begin{eqnarray}
\Gamma_{\mr{eff}} & = & 1-\frac{A}{\Gamma^B}~~{\mr{for}}~~\Gamma > \Gamma_{\mr{crit}},  \nonumber \\
\Gamma_{\mr{eff}} & = & \Gamma~~{\mr{for}}~~\Gamma < \Gamma_\mr{crit}.
\end{eqnarray}
$\Gamma_{\mr{crit}}$ is the critical $\Gamma$ above which inhomogeneities are excited, so the effective opacity for $\Gamma>\Gamma_{\mr{crit}}$ can be written as
\begin{equation}
\kappa_{\mr{eff}}=\kappa_{\mr{Th}}\left(1-\frac{A}{\Gamma^B}\right)/\Gamma.
\label{eq:kappa_eff}
\end{equation}
Since we expect a continuous $\Gamma_{\mr{eff}}$, $A$, $B$ and $\Gamma_{\mr{crit}}$ satisfy the equation 
$\Gamma_{\mr{crit}} = 1- A / \Gamma_{\mr{crit}}^B$.
From theoretical considerations, we take $\Gamma_{\mr{crit}} \sim 0.8$ \citep{ShavivInstabilities}, though we have checked also $\Gamma_{crit}=0.5$, as elaborated in \S\ref{sec:opacityeffect}. This implies\
 a relation between the normalization constant $A$ and the power law $B$, that is given by
\begin{equation}
A=(1-\Gamma_{\mr{crit}}) \Gamma_{\mr{crit}}^B.
\label{eq:A}
\end{equation}
 The other equations describing the structure of this region is the equation of hydrostatic equilibrium, 
\begin{equation}
\frac{dP}{dr}=-g\rho,
\label{eq:hydro_atm}
\end{equation}
where $g=GM/r^2$. To a very good approximation, the mass in the atmosphere is negligible, such that we can safely assume that
\begin{equation}
\frac{dM}{dr}=0.
\label{eq:dm_atm}
\end{equation}

\subsubsection{Continuum Driven Winds}
As we described in \S\ref{sec:SEDwinds}, the average density decreases as we move up the atmosphere. At some point, the typical optical depth of the nonlinear structure would decrease below unity. Once  the typical structure becomes transparent, radiation cannot be funneled around anymore, and the effective opacity returns to its microscopic value. At this point the average radiative force on the gas elements becomes super-Eddington again, and so the gas at this radius experiences a net force outwards. A wind will therefore be accelerated from this radius.

The forces acting on the material in this region are the pressure gradient outwards, and gravity inwards. Thus, the equation of motion of the gas (assuming a steady state wind, namely,  that ${\partial} /{\partial t} \rightarrow 0$) is given by: 
\begin{equation}
v\frac{dv}{dr}=-\frac{GM}{r^{2}}-\frac{1}{\rho}\frac{dP}{dr},
\label{eq:momentum}
\end{equation}
where $v$ is the wind velocity.
The continuity equation implies
\begin{equation}
\dot{m}=4{\pi}r^{2}{\rho}v=const.,
\label{eq:mass_conservetion}
\end{equation}
where $\dot{m}$ is the mass loss rate. 
The equation of energy conservation is
\begin{equation}
L(r)=\dot{E}-\dot{m}\left(\frac{v^{2}}{2}+w(r)-\frac{GM}{r}\right),
\label{eq:energy}
\end{equation}
where $w=5kT/2{\mu}m_H+4P_{rad}/\rho$ is the specific enthalpy and $\dot{E}$ is the total energy loss rate from the sonic point, given by
\begin{equation}
\dot{E}=L_{sonic}+\dot{m}\left[\frac{1}{2}v_{sonic}^{2}+w(R_{sonic})-\frac{GM}{R_{sonic}}\right].
\label{eq:Edot}
\end{equation}
Because the wind is optically thick, the transfer of energy can be described by radiative diffusion. Note that although conditions for convection are still satisfied, we can neglect this energy transport since it is necessarily unimportant in the supersonic flow. Thus, the temperature gradient in the wind is
\begin{equation}
\frac{dT}{dr}=-\frac{3{\kappa}L{\rho}}{16{\pi}acr^{2}T^{3}}.
\label{eq:radiation_transfer}
\end{equation}

\subsubsection{Photon Tired Winds}
The mass loss is predicted according to conditions at the sonic point. If the mass loss is too high and the potential well is too deep, the luminosity may be insufficient to push the wind to infinity. This gives rise to so called photon tired winds \citep{Owocki2004}. 
The mass loss for which the wind becomes photon tired is

\begin{equation}
\dot{m}_\mr{tiring}=\frac{L_\mr{s}R_\mr{s}}{MG}.
\label{eq:photon_tiring}
\end{equation}
The behavoir of photon tired winds was studied by \cite{van_marle2009}. It was found that shocks form between infalling material and outflowing wind. This forms a layer of shocks in which there is a large kinetic flux, but without the associated mass flux. When photon tired winds arise, the mass loss from the top of the layer of shocks is reduced to less than the photon tiring limit, and we take it to be
\begin{equation}
\dot{m}=0.9\dot{m}_\mr{tiring}.
\end{equation} 
\subsubsection{Boundary Conditions}
The structure of the star is determined by the specific entropy at the center (which almost uniquely determines the mass) and by the central density, (see \S\ref{sec:stability}). The only free variable in a given structure is therefore the luminosity. The value chosen for the luminosity should agree with the surface conditions on the radiation field, which should satisfy the blackbody radiation law:
\begin{equation}
L_\mr{ph}=4\pi r_\mr{ph}^2\sigma T_\mr{eff}^4,
\label{eq:bb}
\end{equation} 
where the $T_\mr{eff}$ is the temperature at optical depth $\tau=2/3$. The optical depth in the wind is given by
\begin{equation}
\tau\simeq\kappa\rho r.
\label{eq:optical_depth}
\end{equation}
However, if the wind is sufficiently thick, the gas temperature can decrease enough for recombination of the ionized hydrogen to occur before the photosphere is reached (i.e., at $\tau>1$). This implies that the opacity decreases from its Thomson value to effectively zero, beyond which the gas is transparent. Thus, the photosphere in this case is going to be located at the recombination front. Since it is relatively sharp, we can take the classical Eddington result that $\tau=2/3$ at this radius, and require eq.\ \ref{eq:bb} to be valid there. The actual temperature of recombination depends on the density, but since the structure is not sensitive to the actual recombination temperature, we take a nominal $T_\mr{rec}=4500\mr{K}$.
 As we will show in the results, the states of very low effective temperature, i.e., $T_\mr{eff} \lesssim 4500\mr{K}$ are very common in SMSs. 

\subsection{Evolution of Non-Rotating SMSs}
Assuming that the central temperature never becomes high enough for nuclear burning to be important, and because in the mass range of $4\times10^4$ to $10^8M_{\odot}$ the thermal evolution timescale is longer than the hydrostatic timescale, the evolution of the SMS is described by a sequence of hydrostatic equilibrium states. The star begins as a large spherical cloud with little binding energy. As the star radiates its binding energy, it becomes denser and smaller. In addition, the wind from the star changes the total mass as the star evolves and the central entropy is no longer constant. 
For the evolution of the SMS, we need to know its total energy. The internal component is given by
\begin{equation}
E_\mr{int}=\int_0^R4\pi r^2udr,
\label{eq:Eint}
\end{equation} 
where $u$ is the internal energy density. The gravitational Newtonian energy is given by
\begin{equation}
E_\mr{grav}=-\int_0^R4\pi rGm\rho dr.
\label{eq:Egrav}
\end{equation}
\subsubsection{End Stage of Non-Rotating SMS}
The star ends its hydrostatic equilibrium life when it reaches the critical radius, once  the general relativistic corrections ultimately destabilize the star. This critical radius is given by (see \citealt{ShapiroBook} for details)
\begin{equation}
\frac{GM}{R_\mr{crit}c^2}=0.6295\left(\frac{M_\odot}{M}\right)^{1/2}.
\label{eq:Rcrit}
\end{equation}
 Specifically, the criterion for instability is given in the form of a critical density (eq.~\ref{eq:rho_crit}) above which the destabilizing GR corrections overcome the stabilizing effects of the gas pressure. When this density is reached, the evolution of the star ends and a black hole is formed. 
\subsection{Evolution of Rotating SMSs}
\label{sec:rotating_evolution}
While the initial stages of rotating SMSs are the same as the non-rotating SMSs, the following steps change drastically.
As mentioned in \S\ref{sec:Rotation}, SMSs having enough angular momentum do not collapse when they reach the critical density. Instead, they keep evolving through a sequence of hydrostatic states, while radiating their binding energy and evaporating some of their mass. While contracting, the central temperature increases, and nuclear burning becomes important. At some point, the energy released by nuclear burning becomes large enough to support the SMSs from further contraction. 

We divide the evolution of rotating SMSs into 3 stages:
\begin{itemize}
\item {\em Stage \Rmnum{1}a, Super-Eddington contraction:} The evolution is governed by the SMS contraction, and the radiation of its binding energy. For non-rotating SMSs, this stage ends with GR collapse, which sets the duration of this stage. For rotating SMSs, the stage ends when the central temperature is high enough for hydrogen to be ignited. 
\item {\em Stage \Rmnum{1}b,  Super-Eddington contraction with C buildup:} Although the pp-chain and triple-$\alpha$ burning cannot supply a sufficient amount of energy to halt the contraction, the temperature can be high enough for these reactions to build up $^{12}$C. This build up sets the duration of this stage.
\item {\em Stage \Rmnum{2}, Super-Eddington CNO burning:} The accumulated C slowly increases the fraction of energy produced by the CNO-cycle. Once it becomes dominant, the contraction stops and energy production proceeds through CNO burning. Like the previous stage, this one is super-Eddington and therefore has a continuum driven wind. As a consequence, the evolutionary timescale is governed by mass loss from the atmosphere. As the mass of the SMS decreases, so does its luminosity. 
\item {\em Stage \Rmnum{3}, sub-Eddington, ``normal" massive star:} At this stage, the luminosity is sub-Eddington, and the SMS becomes a ``normal" massive star. Here, the evolutionary timescale is governed by nuclear burning. This final stage is the longest. Since the SMSs at this stage are normal stars, they have already been extensively described in the literature \citep[e.g.,][]{Bond1984}, and therefore will not be described here. 
\end{itemize}
\section{Numerical Methods}
The problem we are required to solve can be divided into two parts. First, we need to solve the hydrostatic model of the SMS, with a given mass and energy. Then, we need to evolve these models to describe the evolution of the SMSs from their formation to their collapse. Here we describe the main methods we used.
\subsection{Solving a Stellar Model}
The stellar structure is obtained by solving eqs.\ \ref{eq:hydrostatic_eq}-\ref{eq:gradT} for the convective region, eqs.\ \ref{eq:dT_atm} and \ref{eq:hydro_atm}-\ref{eq:dm_atm} for the atmosphere and eqs.\ \ref{eq:momentum}-\ref{eq:energy} and \ref{eq:radiation_transfer} for the wind. These equations are solved by first guessing the total luminosity. The solution is then iterated using the shooting method, until the outer boundary condition given in eq.\ \ref{eq:bb} is fulfilled.
\subsection{Evolving the Stellar Models}
After we obtain a single snapshot of a SMS, we can proceed to find how it evolves with time. 
Each model solution predicts a luminosity $L$ and a mass loss rate $\dot{m}$. The stellar mass and total energy are then evolved through conservation of mass and energy:
\begin{equation}
\frac{dM}{dt}=-\dot{m}
\label{eq:dM_dt}
\end{equation} 
\begin{equation}
\frac{dE}{dt}=L_\mathrm{nuc}-L,
\label{eq:dE_dt}
\end{equation}
where $L_\mathrm{nuc}$ is the total nuclear energy generation rate, while $L$ is the total luminosity. $E$ is the sum of the internal and gravitational energies. 
 
Since $\rho_c$ and $s$ are the real variables describing the models, and not $M$ and $E$, the model evolution proceeds by choosing a small central density increment $\Delta \rho_c = \epsilon \rho_c$ (specifically with $\epsilon = 0.001$), and then solving for the new model.   
 	
\section{Results}
We now proceed to describe the results of our simulations. We present the structure and evolution of the SMSs from their early phase to the time they reach the critical radius and collapse, if they are non-rotating, or until they become normal massive stars, if they do rotate. 

In particular, we perform a parameter study. Since the exact ``porous'' opacity law required is not adequately known, we depict solutions of the same mass, but different effective opacity laws. This allows us to study the influence of the atmospheric effective opacity on the stellar structure and evolution.
\subsection{Quasistatic Configuration}
\begin{figure}
\centerline{\epsfig{file=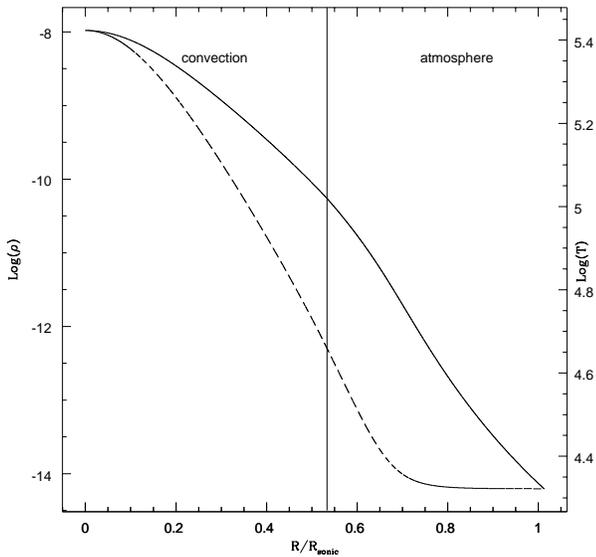,width=8cm,angle=0}}
\caption{Gas density $\rho(r)$ (solid) and temperature (dashed) vs. stellar radius at the convective zone and the atmosphere for a SMS with $5.5\times10^4M_\odot$, in its early phase ($R_\mr{sonic}\approx2000R_\mr{crit}\approx6\times10^{15}cm$).}
\label{fig:snap1}
\end{figure}

\begin{figure}
\centerline{\epsfig{file=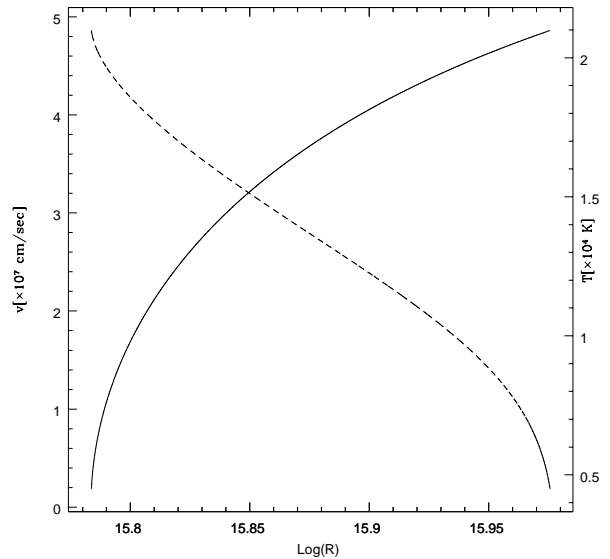,width=8cm,angle=0}}
\caption{The wind velocity (solid) and the wind temperature (dashed) vs.\  radius for the same object as fig.~\ref{fig:snap1}.}
\label{fig:snap2}
\end{figure}

The early phase ($R_\mr{sonic}\approx2000R_\mr{crit}$) of a SMS with $M_0\approx5.5 \times 10^4 M_\odot$ is depicted in figs.\ \ref{fig:snap1}-\ref{fig:snap2}. In fig.\ \ref{fig:snap1} we present the gas density and the temperature in both the convective zone and the atmosphere. In the latter, the density decreases by three orders of magnitude, enabling a much ``lighter'' wind.

The wind velocity and temperature are described in fig.\ \ref{fig:snap2}, from the sonic point to the photosphere. In the case depicted here, the wind is optically and geometrically thick ($R_\mr{ph}\sim1.5R_\mr{sonic}$).

\subsection{Stage Ia: Temporal Evolution Before Collapse or Ignition of Nuclear Burning}
\begin{figure}
\centerline{\epsfig{file=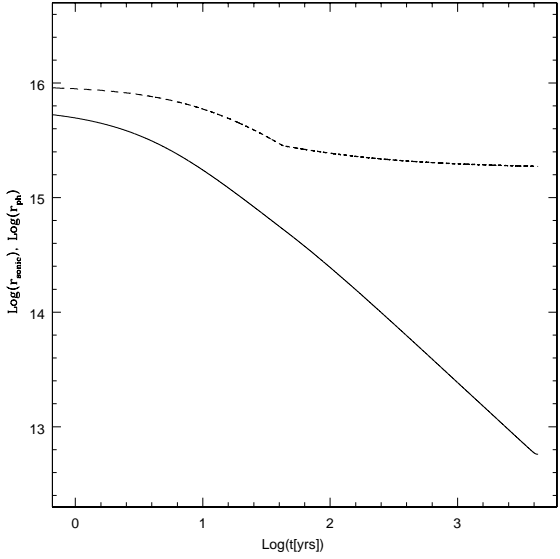,width=8cm,angle=0}}
\caption{The sonic radius $R_\mr{sonic}$  (solid), and the radius of the photosphere $R_\mr{ph}$ (dashed), as function of time for a star with an initial mass of $M=5.5\times10^4M_\odot$. }
\label{fig:evolution1_1e38}
\end{figure}

\begin{figure}
\centerline{\epsfig{file=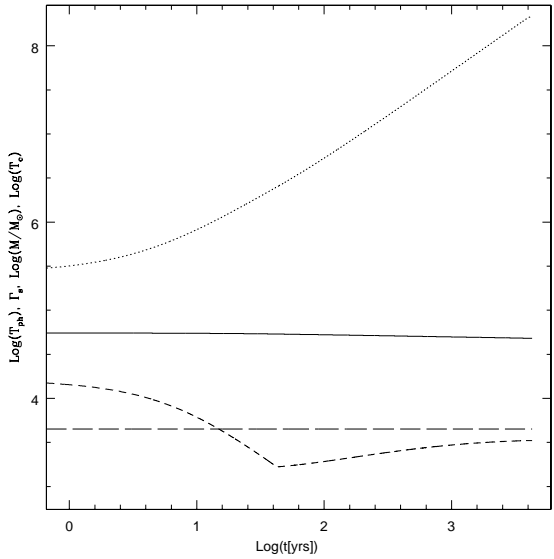,width=8cm,angle=0}}
\caption{The evolution of the mass (solid), luminosity at the base of the wind $L_\mr{s}/L_\mr{Edd}$ (dashed), central temperature $T_c$ (dotted) and effective temperature $T_\mr{ph}$ (long dashed) as a function of time, for a star with an initial mass of $M=5.5\times10^4M_\odot$. The SMS loses about 13$\%$ of its mass until ingnition of hydrogen. Note that $T_\mr{ph}$ stays low during this stage, so no ionizing radiation is emitted.}
\label{fig:evolution2_1e38}
\end{figure}

Figs.\ \ref{fig:evolution1_1e38}-\ref{fig:evolution2_1e38} plot the temporal evolution of the sonic and photospheric radii, the luminosity and the mass of a star with an initial mass of $M=5.5\times10^4M_\odot$. This star loses about 13\% of its initial mass during this stage (until ingnition of nuclear burning). As can be seen in fig.\ \ref{fig:evolution1_1e38}, the wind is optically thick, causing the effective temperature to remain low.  Thus, a negligible amount of ionizing radiation is emitted during this stage of evolution, unlike the standard lore for SMSs.   

The results for non-rotating SMSs (which only have this stage of evolution) are summarized in table \ref{tab:results}, for different initial masses.

\subsection{Ignition and Evolution of Rotating SMSs}
\begin{figure}
\centerline{\epsfig{file=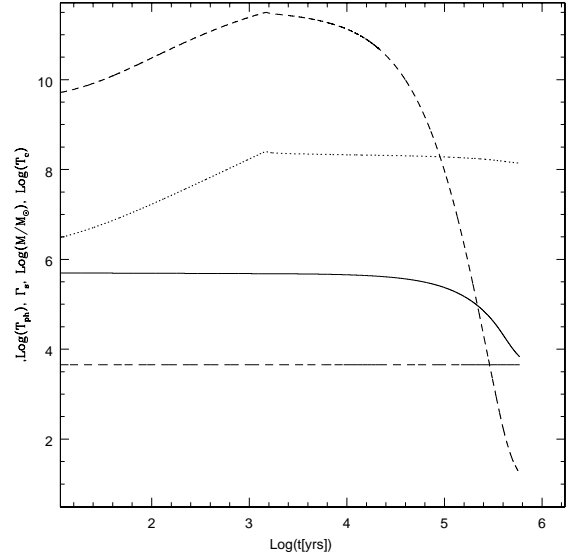,width=8cm,angle=0}}
\caption{The evolution (stages \Rmnum{1}b-\Rmnum{2}) of a rotating star with an initial mass $M=5\times10^5 M_{\odot}$. Plotted are the mass (solid), luminosity at the base of the wind ($L_s/L_{Edd}=\Gamma_s$, dashed), central temperature (dotted) and effective temperature (long dashed). The effective temperature remains low during all stages of evolution, this SMS emitts very small amount of ionizing radiation.}
\label{fig:Rot2}
\end{figure}

\begin{figure}
\centerline{\epsfig{file=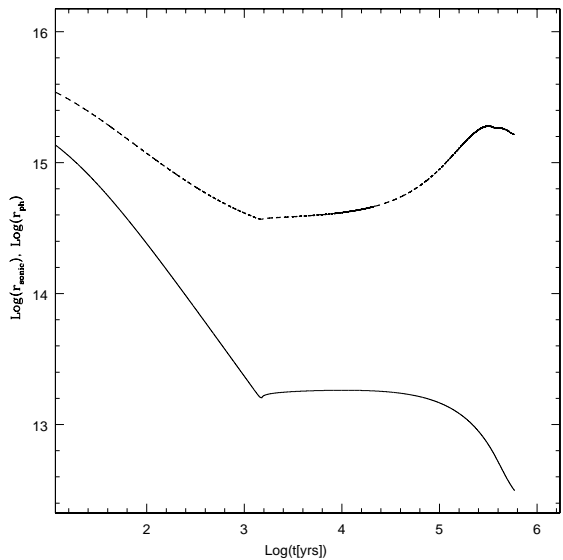,width=8cm,angle=0}}
\caption{Sonic radius $r_{sonic}$ (solid), and radius of the photosphere (dashed) of the same SMS as fig.\ \ref{fig:Rot2}, during stages Ib and II of the evolution. In stage III (not shown), the luminosity is sub-Eddington, so there is no continuum driven wind and the photosphere is located at the top of the atmosphere.}
\label{fig:Rot1}
\end{figure}

Figs.~\ref{fig:Rot2}-\ref{fig:Rot1} show stages Ib and II of a rotating SMS with an initial mass $M=5\times10^{5}M_\odot$. Specifically, fig.\  \ref{fig:Rot2} depicts the mass, luminosity (as $\Gamma_s$), central temperature ($T_\mr{c}$) and effective temperature ($T_\mr{ph}$) of this SMS. Fig.\ \ref{fig:Rot1} describes the sonic and photospheric radii as a function of time. The two stages Ib and II discussed in \S\ref{sec:rotating_evolution} are easily distinguished. Stages Ib lasts $\sim 10^3$yr. Here the SMS radiates its binding energy, while contracting (as can be seen as a fast decrease of the sonic radius). The central temperature increases from $\sim 10^6$ to $>10^8$ and the luminosity is very high ($>10L_{Edd}$). Stage II lasts for $\sim 10^6$yr. Here the evolution is ruled by the wind mass loss from the atmosphere. The total mass and luminosity decrease, but the central temperature and the sonic radius remain almost constant. Stage III (not shown) is the last and longest. At this stage, the luminosity is sub-Eddington and the evolution is ruled by the nuclear reactions. If a wind is present, it will be a weak line driven wind due to the small amount of CNO that was produced.

\subsection{The effects of the atmospheric opacity law}
\label{sec:opacityeffect}
One of the uncertainties in the model, is the opacity law behavior of the porous atmosphere. The reasons it is not known well is because it depends on the nonlinear radiative hydrodynamic configuration the atmosphere will reach, and unfortunately, there are still no numerical simulations or empirical data which can constrain the effective opacity law. It is for this reason that we parameterized the effective opacity using eq.\ \ref{eq:kappa_eff}, with one free parameter, $B$. We have examined the influence of different values of $B$ on the stellar structure and evolution. The values for $B$ taken were $B=0.75$, $B=1.0$ and $B=1.25$.

We find that there is no significant difference in the appearance of SMSs, under different opacity laws. However, there are some differences with respect to SMS evolution.

Because the luminosity depends on the opacity, the lifetime somewhat depends on the opacity law. For example, the lifetime of stages Ia and Ib of a $5.5 \times 10^4 M_\odot$ SMS (i.e., the lifetime of the gravitational collapse until ignition of CNO nuclear burning), ranges between 2800 yr for $B=0.75$ and 7000 yr for $B=1.25$.

However, there is one major characteristic which is sensitive to the opacity law, and it is the mass of the sub-Eddington star that will end the super-Eddington phase. The opacity law can be described by a $\Gamma_{crit}$, which is the Eddington parameter below which the system becomes sub-Eddington. Using the Eddington quartic relation and the approximation that the SMS can be described by an $n=3$ polytrope, it is possible to estimate the critical Mass for which the SMS will become sub-Eddington \citep[see][]{OwockiShavivChapter}:
\begin{equation}
M_{crit} = {\mu^{-2} \Gamma_{crit}^{1/2} \over (1-\Gamma_{crit})^2}18.3 M_\odot,
\end{equation} 
where $\mu$ is the molecular weight. 

Since there is little depletion of hydrogen during the SMS evolution, we can take the primordial value of $X\approx 0.75$. Moreover, \cite{ShavivInstabilities} found two instabilities, one with $\Gamma_{crit}\approx 0.8$, which is more general, and one with $\Gamma_{crit}=0.5$, which depends on the boundary conditions. Other instabilities, when magnetic fields are present could operate at even lower values of $\Gamma_{crit}$ \cite[see for example the numerical analysis of instabilities in luminous accretion disks by][]{Turner2005}. For the above value of $X$ and range of $\Gamma_{crit}$, we find $M_{crit}$ ranging between 150 and 1200 $M_\odot$. Namely, the ``normal" sub-Eddington star left once the super-Eddington phase is over is very sensitive to the critical Eddington parameter above which atmospheres become porous.





\begin{table*}
\begin{tabular}{c c c c c c c}
\hline
$M_\mr{initial}$&$\tau_{collapse}(lifetime)$&$M(\tau)/M_\mr{inital}$&$T_c({\tau})$&${\rho}_c({\tau})$\\
$\left[ M_\odot \right]$ & $\left[yr\right]$ & $$ & $\left[ K \right]$ &  $\left[ gr~cm^{-3} \right]$ \\
\hline \hline
$5.5\times10^4$&$--$&$0.87$&$2.1\times10^8$&$6$\\
\hline

$1.0\times10^5$&$--$&$0.86$&$2.2\times10^8$&$4.9$\\
\hline
$2.5\times10^5$&$8.6\times10^2$&$0.94$&$1.0\times10^8$&$0.3$\\
\hline
$5\times10^5$&$3.2\times10^2$&$0.97$&$5.2\times10^7$&$2.7\times10^{-2}$\\
\hline
$1.5\times10^6$&$60$&$0.98$&$1.7\times10^7$&$5\times10^{-4}$\\
\hline
$5.0\times10^6$&$10$&$>0.99$&$5.0\times10^6$&$8\times10^{-6}$\\
\hline
$1.0\times10^7$&$<10$&$1.00$&$2.5\times10^6$&$7.5\times10^{-7}$\\

\end{tabular}
\caption{Summary of the pre-collapse stage of non-rotating SMSs . Tabulated are the various SMS characteristics as a function of the initial mass. A few points to note: (A) The two lower masses represented ($5.5\times10^4M_\odot$ and $1.0\times10^5M_\odot$) will not collapse even if there is no rotation, because at some point of their evolution the thermonuclear energy generation becomes important, supporting the star against additional contraction. Thus, these objects are not SMSs, but ``Very Massive Stars" instead (e.g., see \citealt{Bond1984}) (B) The SMSs always have a large enough photosphere, hence,  a low effective temperature (determined by recombination in the wind). As a consequence, SMSs will not contribute a significant amount of ionizing radiation. (C) The mass shed by the objects is small relative to the total mass of the star. Thus, as long as thermonuclear reactions are not ignited, the total wind mass loss does not significantly affect the evolution. (D) Masses larger than the above range may collapse to become a black hole, but they do not form an intermediate hydrostatic object because of the short gravitational collapse time scale. }
\label{tab:results}
\end{table*}

\section{Summary \& Discussion}
We obtained numerical solutions for rotating and non-rotating SMSs. In the range of masses $5\times10^4$ to $1\times10^7M_\odot$. In this range, the hydrodynamical timescale is shorter than the thermal timescale, such that the star is always in hydrostatic equilibrium during its evolution. Above these masses, there is no equilibrium phase and the star collapses on a dynamical timescale. 
For SMSs with $M>1\times10^5M_{\odot}$, the thermonuclear energy generation of the non-rotating SMSs is negligible, thus, in this range, the SMSs are  hydrostatic objects which slowly release their binding energy, and shrink until they end their life through a collapse into a super massive black hole. 
The rotating SMSs will not collapse, but instead ignite nuclear burning and lose mass  through a wind, until a sub-Eddington star is left. The mass of this star depends on the Eddington parameter $\Gamma_{crit}$ for which the system becomes sub-Eddington (see \S\ref{sec:opacityeffect}).  
\\ \\
 Our analysis included two main parts:
\begin{itemize}
\item We constructed hydrostatic models of SMSs. These included the full calculation of the SMS structure while considering that luminous  atmospheres are unstable and become ``porous'' as the Eddington luminosity is approached, and that continuum driven winds are accelerated by these objects. When relevant, it also included nuclear reactions.

\item We then evolved the models for SMSs using the hydrostatic configurations as basic building blocks. The evolution allows us to calculate various general properties such as the total mass loss, the energy radiated in ionizing photons, or the total life time. 
\end{itemize} 
The main results obtained are the following:
\begin{itemize}
\item The polytropic solution is a very good approximation where convection is dominant, but a very poor approximation in the outer parts of the star. We also note that some formation scenarios can produce a significantly non-polytropic structure \citep{Begelman2010}, but these scenarios were not discussed here. 
\item The solution describes objects which radiate above their Eddington limit, in contrast to the standard solution where the luminosity is just below the Eddington limit. The luminosities obtained vary between $\sim 2\ledd$ to $\sim 50\ledd$ at the base of the wind (the larger the mass the higher the luminosity), but the actual luminosities emitted from the photosphere vary between $\sim10^{-3}\ledd$ to $\sim2\ledd$ (the larger the mass, the lower is the photospheric luminosity). The rest of the energy is used to drive a wind. 
\item Because the binding energy at the onset of the GR collapse into a black hole does not depend on the outer layers of the star, higher luminosities imply shorter lifetimes in non-rotating stars. 
\item The mass loss driven by the wind in the pre-collapse phase, is typically between $5\%$ to $20\%$ of the total mass, where less massive SMSs have larger mass loss. This immediately implies mass loss will be crucial in stars less massive than the least massive SMSs. These stars, which have prolonged lifetimes (due to nuclear reactions), will lose significant amounts of mass, making mass loss the dominant factor in their evolution.  
\item The effective temperature of the SMSs remains low during their evolution (i.e., the photosphere is determined by recombination at the wind, taking place at $\sim 4500\mr{K}$) due to the large photospheric radius. Hence, the amount of ionizing radiation emitted by these objects is negligible.
\item Rotating SMSs will lose most of their mass in the later stage (\Rmnum{2}, see \S\ref{sec:rotating_evolution}), through their winds. The mass loss continues until a sub-Eddington massive star is left.  
\end{itemize}

\subsection{Differences from the Standard Solution}
As mentioned before, the solution for SMSs obtained here is notably different from the standard lore of SMSs. It is thus worth while to emphasize these differences and their causes.

In SMSs in general, the radiation to gas pressure ratio is very large. If the opacity is not modified and instead remains at its microscopic value of Thomson scattering, it is straightforward to show that the luminosity should be that of Eddington minus a small relative correction of order the gas to total pressure ratio. Hence,
the luminosity in the classical SMS solutions is taken to be the Eddington luminosity.

However, this does not consider the rise of instabilities when the radiation pressure is dominant. Once taken into consideration, the effective opacity is reduced and the effective Eddington luminosity is increased. This gives rise to solutions with $L>\ledd$, in which the outer regions of the hydrostatic part of the star is porous.

Since the structure of the convective zone does not depend on the luminosity, this inner region (which comprises most of the stellar mass), does not change in the super Eddington solution. Nevertheless, this region does not fit exactly the polytropic solution with a polytropic index of $n=3$. This deviation from the polytropic solution is caused by two reasons: the finite pressure contribution by the gas and the limited efficiency of convection at its upper boundary, both effects are not due to the super-Eddington nature of the solution.

Another intrinsic characteristic of the present solutions is the high mass loss rate. This has several effects. First,  the high mass loss rates imply that large amounts of energy are required to push the gas out of the gravitational potential well. As a result, the luminosity left above the wind can be significantly reduced, and even make the intrinsically super-Eddington SMSs appear to be sub-Eddington.

Second, it implies that the photosphere resides at large radii, inside the wind.
As a consequence, the photospheric temperature is lower by an order of magnitude relative to the standard SMS solution. As a consequence, only a negligible amount of ionizing radiation is emitted. This is in markable contrast with the standard SMS models, and it implies that it is unlikely that SMSs have played a major role in reionizating the early universe.

\section*{Acknowledgements}

This research has been supported by the Israel Science Foundation, grant 1589/10.

\bibliographystyle{mn2e}
\bibliography{thebibfile}{}

\begin{thebibliography}{}

\bibitem[\protect\citeauthoryear{{Arons}}{{Arons}}{1992}]{Bubbles1}
{Arons} J.,  1992, \apj, 388, 561

\bibitem[\protect\citeauthoryear{{Begelman}}{{Begelman}}{2002}]{Bubbles3}
{Begelman} M.~C.,  2002, \apjl, 568, L97

\bibitem[\protect\citeauthoryear{{Begelman}}{{Begelman}}{2010}]{Begelman2010}
{Begelman} M.~C.,  2010, \mnras, 402, 673

\bibitem[\protect\citeauthoryear{{Bond}, {Arnett} \& {Carr}}{{Bond}
  et~al.}{1984}]{Bond1984}
{Bond} J.~R.,  {Arnett} W.~D.,    {Carr} B.~J.,  1984, \apj, 280, 825

\bibitem[\protect\citeauthoryear{{Dotan}, {Rossi} \& {Shaviv}}{{Dotan}
  et~al.}{2011}]{DotanHalos}
{Dotan} C.,  {Rossi} E.~M.,    {Shaviv} N.~J.,  2011, \mnras, 417, 3035

\bibitem[\protect\citeauthoryear{{Dotan} \& {Shaviv}}{{Dotan} \&
  {Shaviv}}{2011}]{DotanDisks}
{Dotan} C.,  {Shaviv} N.~J.,  2011, \mnras, 413, 1623

\bibitem[\protect\citeauthoryear{{Fowler}}{{Fowler}}{1966}]{Fowler66}
{Fowler} W.~A.,  1966, \apj, 144, 180

\bibitem[\protect\citeauthoryear{{Gammie}}{{Gammie}}{1998}]{Bubbles2}
{Gammie} C.~F.,  1998, \mnras, 297, 929

\bibitem[\protect\citeauthoryear{{Glatzel}}{{Glatzel}}{1994}]{Opacity1}
{Glatzel} W.,  1994, \mnras, 271, 66

\bibitem[\protect\citeauthoryear{{Joss}, {Salpeter} \& {Ostriker}}{{Joss}
  et~al.}{1973}]{Joss1973}
{Joss} P.~C.,  {Salpeter} E.~E.,    {Ostriker} J.~P.,  1973, \apj, 181, 429

\bibitem[\protect\citeauthoryear{{Owocki} \& {Gayley}}{{Owocki} \&
  {Gayley}}{1997}]{PhotonTired}
{Owocki} S.~P.,  {Gayley} K.~G.,  1997, in ASP Conf. Ser. 120: Luminous Blue
  Variables: Massive Stars in Transition {ThePhysics of Stellar Winds Near the
  Eddington Limit}.
p.~121

\bibitem[\protect\citeauthoryear{{Owocki}, {Gayley} \& {Shaviv}}{{Owocki}
  et~al.}{2004}]{Owocki2004}
{Owocki} S.~P.,  {Gayley} K.~G.,    {Shaviv} N.~J.,  2004, \apj, 616, 525

\bibitem[\protect\citeauthoryear{{Owocki} \& {Shaviv}}{{Owocki} \&
  {Shaviv}}{2012}]{OwockiShavivChapter}
{Owocki} S.~P.,  {Shaviv} N.~J.,  2012, in {Davidson} K.,  {Humphreys} R.~M.,
  eds, , Eta Car and the SN Impostors.
Springer

\bibitem[\protect\citeauthoryear{{Papaloizou}, {Alberts}, {Pringle} \&
  {Savonije}}{{Papaloizou} et~al.}{1997}]{Opacity2}
{Papaloizou} J.~C.~B.,  {Alberts} F.,  {Pringle} J.~E.,    {Savonije} G.~J.,
  1997, \mnras, 284, 821

\bibitem[\protect\citeauthoryear{{Prialnik} \& {Kovetz}}{{Prialnik} \&
  {Kovetz}}{1992}]{Prialnik1992}
{Prialnik} D.,  {Kovetz} A.,  1992, \apj, 385, 665

\bibitem[\protect\citeauthoryear{{Shapiro} \& {Teukolsky}}{{Shapiro} \&
  {Teukolsky}}{1983}]{ShapiroBook}
{Shapiro} S.~L.,  {Teukolsky} S.~A.,  1983, {Black holes, white dwarfs, and
  neutron stars: The physics of compact objects}.
Research supported by the National Science Foundation.~New York,
  Wiley-Interscience, 1983, 663 p.

\bibitem[\protect\citeauthoryear{{Shaviv}}{{Shaviv}}{1998}]{ShavivPorous}
{Shaviv} N.~J.,  1998, \apjl, 494, L193

\bibitem[\protect\citeauthoryear{{Shaviv}}{{Shaviv}}{2000}]{ShavivEta}
{Shaviv} N.~J.,  2000, \apjl, 532, L137

\bibitem[\protect\citeauthoryear{{Shaviv}}{{Shaviv}}{2001a}]{ShavivInstabiliti%
es}
{Shaviv} N.~J.,  2001a, \apj, 549, 1093

\bibitem[\protect\citeauthoryear{{Shaviv}}{{Shaviv}}{2001b}]{ShavivNovae}
{Shaviv} N.~J.,  2001b, \mnras, 326, 126

\bibitem[\protect\citeauthoryear{{Turner}, {Blaes}, {Socrates}, {Begelman} \&
  {Davis}}{{Turner} et~al.}{2005}]{Turner2005}
{Turner} N.~J.,  {Blaes} O.~M.,  {Socrates} A.,  {Begelman} M.~C.,    {Davis}
  S.~W.,  2005, \apj, 624, 267

\bibitem[\protect\citeauthoryear{{van Marle}, {Owocki} \& {Shaviv}}{{van Marle}
  et~al.}{2009}]{van_marle2009}
{van Marle} A.~J.,  {Owocki} S.~P.,    {Shaviv} N.~J.,  2009, \mnras, 394, 595

\bibitem[\protect\citeauthoryear{{Wagoner}}{{Wagoner}}{1969}]{Wagoner1969}
{Wagoner} R.~V.,  1969, \araa, 7, 553

\bibitem[\protect\citeauthoryear{{Weinberg}}{{Weinberg}}{1972}]{WeinbergBook}
{Weinberg} S.,  1972, {Gravitation and cosmology: Principles and applications
  of the general theory of relativity}.
New York: Wiley, |c1972

\end{thebibliography}

\end{document}